\documentclass[12pt,manuscript]{aastex}

\usepackage{natbib}

\usepackage{latexsym}
\usepackage{graphics}
\usepackage{graphicx}
\usepackage{natbib}
\usepackage{amsmath}
\usepackage{subfigure}
\usepackage[english]{babel}
\usepackage{makeidx}
\usepackage{amssymb} %tipos de simbolos

\newcommand{\II}{\scriptsize{II}\normalsize}

\shorttitle{Activity cycles in dM stars}
\shortauthors{Buccino et al.}

\begin{document}


\title{Long-term chromospheric activity in southern M dwarfs: \\
Gl 229 A and Gl 752 A}


\author{Andrea P. Buccino\altaffilmark{1}}
\affil{Instituto de Astronom\'\i a y F\'\i sica del Espacio (CONICET),  C.C. 67 Sucursal 28, C1428EHA-Buenos Aires Argentina}
\affil{Departamento de F\'\i sica. Facultad de Ciencias Exactas y Naturales. Universidad de Buenos Aires}
\email{abuccino@iafe.uba.ar}

\author{Rodrigo F. D\'\i az\altaffilmark{1}}
\affil{Institut d'Astrophysique de Paris, 98bis, bd Arago - F-75014 Paris France}
\author{Mar\'\i a Luisa Luoni}
\affil{Instituto de Astronom\'\i a y F\'\i sica del Espacio (CONICET),  C.C. 67 Sucursal 28, C1428EHA-Buenos Aires Argentina}
\author{Ximena Abrevaya}

\affil{Instituto de Astronom\'\i a y F\'\i sica del Espacio (CONICET),  C.C. 67 Sucursal 28, C1428EHA-Buenos Aires Argentina}

\and
\author{Pablo J. D. Mauas\altaffilmark{1}}

\affil{Instituto de Astronom\'\i a y F\'\i sica del Espacio (CONICET),  C.C. 67 Sucursal 28, C1428EHA-Buenos Aires Argentina}

\altaffiltext{1}{Visiting Astronomer, Complejo Astron\'omico El
Leoncito operated under agreement between the Consejo Nacional de
Investigaciones Cient\'\i ficas y T\'ecnicas de la Rep\'ublica
Argentina and the National Universities of La Plata, C\'ordoba and San
Juan.}

\begin{abstract}
Several late-type stars present activity cycles similar to that of the Sun.  However, these cycles have been mostly studied in F to K stars.
Due to their small intrinsic brightness, M dwarfs are not
usually the targets of long-term observational studies of stellar
activity, and their long-term variability is generally not
known. In this work, we study the long-term activity of two M dwarf
stars: Gl 229 A (M1/2) and Gl 752 A (M2.5).  We employ medium
resolution echelle spectra obtained at the 2.15 m telescope at the
Argentinian observatory CASLEO between the years 2000 and
2010 and photometric observations obtained from the ASAS database. We
analyzed Ca \II\- K line-core fluxes and the mean V magnitude with the
Lomb-Scargle periodogram, and we obtain possible activity cycles of
$\sim$4 yr and $\sim$7 yr for Gl 229 A and Gl 752 A respectively.

\end{abstract}





\section{Introduction}

During the last years, a new interest on M dwarfs has emerged. One
reason is because their low masses make them ideal targets around
which to search for terrestrial planets in the habitable zone
(e.g. \citealt{2009Natur.462..891C}). However, these stars can be very
active and their activity signatures can hinder the detection of
orbiting planets (e.g. \citealt{2007A&A...474..293B}). Furthermore,
the levels of UV radiation, which are strongly related to stellar
activity, can also limit the habitability
\citep{2006Icar..183..491B,2007Icar..192..582B}.
%These facts make the analysis of activity in M stars relevant.

The usually accepted model to describe the generation and
intensification of magnetic fields in late F- to early M-type stars is
the $\alpha\Omega$-dynamo first invoked to explain solar activity
(\citealt{1955ApJ...122..293P}). It is this model, where the
large-scale magnetic field generation results from the interaction of
differential rotation in the tachocline and the convective turbulence,
which predicts a strong correlation between activity and rotation.
Magnetic activity therefore decays with time as the star spins down
due to braking by magnetized winds. However, recent
 results show that
this decay happens in the first two Gyr of a star's life, to be
compared with lifetimes of late-type stars which are at least 2 times
larger \citep{2004A&A...426.1021P}.

Since stars with spectral type later than M3 are believed to be fully
convective, they should not be able to sustain a solar-like
$\alpha\Omega$ dynamo. Furthermore, most  dM stars are usually slow
rotators. In a recent survey of 123 M-dwarfs, 
\cite{2010AJ....139..504B} found that only seven of them are rotating more
rapidly than their detection threshold of $v\sin i\approx 2.5\,\,\rm{km\,\,
s}^{-1}$. However, 
there is plenty of observational evidence that late-type slow rotators
like dMe stars are very active and have strong magnetic fields \citep{1996ApJ...459L..95J}, with filling factors larger
than for earlier stars
\citep{1980ApJ...239L..27M,1991ApJ...378..725H, 1994A&A...281..129M,
1996A&A...310..245M, 1997A&A...326..249M}. In fact, there is evidence
that stellar
spin-down and active lifetimes change near the mass where full
convection sets in \citep{1998A&A...331..581D,2003ApJ...586..464B,2008ApJ...684.1390R,2008AJ....135..785W}.

\cite{2006A&A...446.1027C} 
proposed that for fully convective dM stars, large-scale magnetic
fields could be produced by a pure $\alpha^2$ dynamo, where activity
would not decay with time since it does not involve
rotation. \cite{2008ApJ...676.1262B} developed a 3-D dynamo model for M dwarfs, 
and found that fully convective stars can generate kG-strength
magnetic fields without the aid of a shearing tachocline.

 To date, activity cycles have been detected in several late-type
stars
(e.g. \citealt{1995ApJ...438..269B,2008A&A...483..903B}). However,
these works are mainly concentrated on F to K stars, since due to the
long exposure times needed to observe them, the red and faint M dwarfs
are usually not the target of long-term observational studies of
stellar activity. As a contribution to this subject, we have developed
an observing program at the Argentinian observatory \emph{Complejo
Astronomico El Leoncito} (CASLEO) dedicated to periodically obtain
medium resolution echelle spectra of southern late-type stars,
including those stars which are fully-convective. Our program has been
operating since 1999. From our data, we found evidences of cyclic
activity for the dMe 5.5 star Prox Centauri, with P = 442 $\pm$ 45
days \citep{2007A&A...461.1107C}, and for the dMe 3.5 spectroscopic
binary Gl 375, with P=763 $\pm$ 40 days \citep{2007A&A...474..345D}.

In this work we study the long-term activity of the dM stars Gl 229 A
and Gl 752 A.  In section \S\ref{sec.obs} we describe the
spectroscopic observations obtained and the photometric data used to
analyze these stars. In section \S\ref{sec.gj752} and
\S\ref{sec.gj229} we report our results. In section \S\ref{sec.halfa}
we analyze the role of H$\alpha$ as an activity indicator in these
stars.

\section{Observations}\label{sec.obs}

%% In a manner similar to \objectname authors can provide links to dataset
%% hosted at participating data centers via the \dataset{} command.  The
%% second curly bracket argument is printed in the text while the first
%% parentheses argument serves as the valid data set identifier.  Large
%% lists of data set are best provided in a table (see Table 3 for an example).
%% Valid data set identifiers should be obtained from the data center that
%% is currently hosting the data.
%%
%% Note that AASTeX interprets everything between the curly braces in the 
%% macro as regular text, so any special characters, e.g. "#" or "_," must be 
%% preceded by a backslash. Otherwise, you will get a LaTeX error when you 
%% compile your manuscript.  Special characters do not 
%% need to be escaped in the optional, square-bracket argument.

Since 1999, we systematically observed more than 140 main-sequence
stars from F5.5 to M5.5, with the echelle spectrograph on the 2.15 m
telescope of the CASLEO Observatory located in the Argentinean
Andes. To date, we have more than 4000 spectra, ranging from 3890 to
6690 \AA\- with R =$\lambda/\Delta\lambda\approx$26400, which
constitute an ideal dataset to study long-term activity. These spectra
are calibrated in flux according to the method outlined in
\cite{2004A&A...414..699C}, and allow us to simultaneously study
different spectral features, from the Ca \II\- lines to H$\alpha$
(e.g. \citealt{2007MNRAS.378.1007D}).

The usual indicator of chromospheric activity in dF to dK
stars is the well known Mount Wilson $S$ index, essentially the ratio
of the flux in the core of the Ca \II\- H and K lines to the continuum
nearby \citep{1978PASP...90..267V}.  \cite{2007astro.ph..3511C}
defined a Ca \II\- index from the CASLEO spectra, which was calibrated
to the Mount Wilson index. Since CASLEO spectra are calibrated in
flux, they also derived a conversion factor which translates known $S$
index to flux in the Ca \II\- cores.  However, due to the low intrinsic
luminosity of M dwarfs, the low signal-to-noise in the Ca
\II\- H line in several CASLEO spectra and the fact that this line can
be contaminated by the H$\epsilon$ line, the $S$ index is not suitable
to study the chromospheric activity on the faint stars in
this study. In this paper we use as a proxy of
stellar activity the Ca \II\- K line-core flux, integrated with a
triangular profile of 1.09 \AA\- FWHM to mimic the response of the
Mount Wilson instrument.

We complement our data with photometry from the All Sky Automated
Survey (ASAS). This project is dedicated to constant photometric
monitoring of the whole available sky, including approximately 10$^7$
stars brighter than 14 magnitude
\citep{2002AcA....52..397P}. Presently, ASAS consists of two observing
stations, one in Las Campanas Observatory, Chile (since 1997) and the
other on Haleakala, Maui (since 2006). Each site is
equipped with two wide-field 200/2.8 instruments, observing
simultaneously in the V and I bands\footnote{The public data are
available at \textsf{http://www.astrouw.edu.pl/asas/}.}. In this work
we used the mean V magnitude of each observing season to analyze the long-term activity
of the stars. 

%% In this section, we use  the \subsection command to set off
%% a subsection.  \footnote is used to insert a footnote to the text.


%% This section also includes several of the displayed math environments
%% mentioned in the Author Guide.

\section{Gl 229 A - HD 42581}\label{sec.gj229}

\begin{figure}[htb!]
\centering
\includegraphics[width=.7\textwidth]{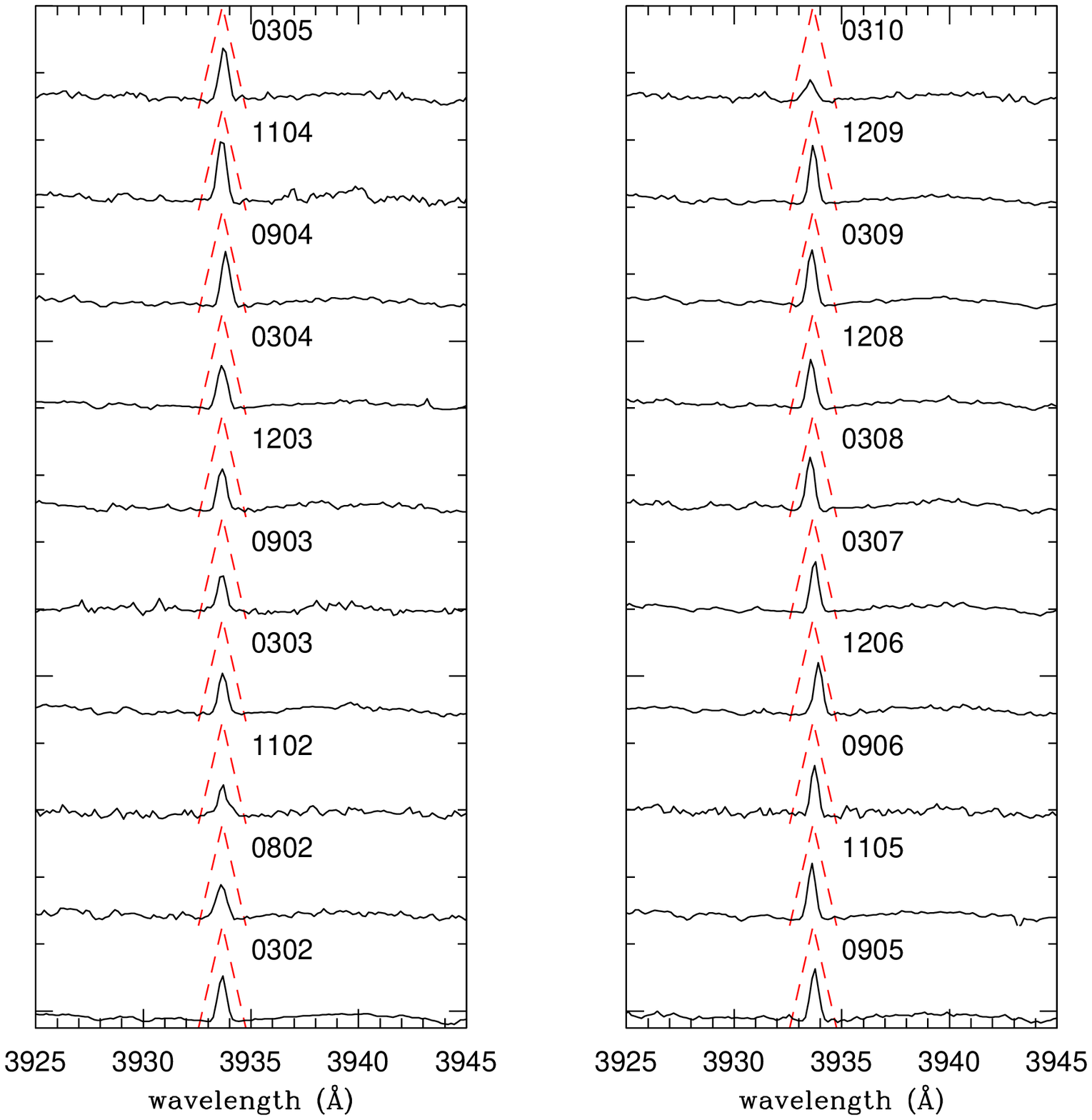}
\caption{Gl 229 A. Ca \II\- K-line in the CASLEO spectra. The triangular profile used to integrate the fluxes is also indicated with dashed line.}\label{fig.sp_hd42581}
\end{figure}

Gl 229 A is a moderately active dM1 flare star
\citep{1985MNRAS.214..119B}, with a brown dwarf companion \citep{1995Natur.378..463N}. We  observed Gl 229 A for 20 nights between the years 2002 and
2010.  In Fig.~\ref{fig.sp_hd42581} we plot each spectrum, with a 
label which indicates the month and year of the observation (e.g. 0302 means
that the observation was obtained in March, 2002).  
We excluded the
spectrum obtained in November 2004 due to its low signal-to-noise near
the Ca \II\- lines and the one observed in March 2010 because it was 
 obtained in a cloudy night.

\begin{figure}[htb!]
\centering
\includegraphics[width=.7\textwidth]{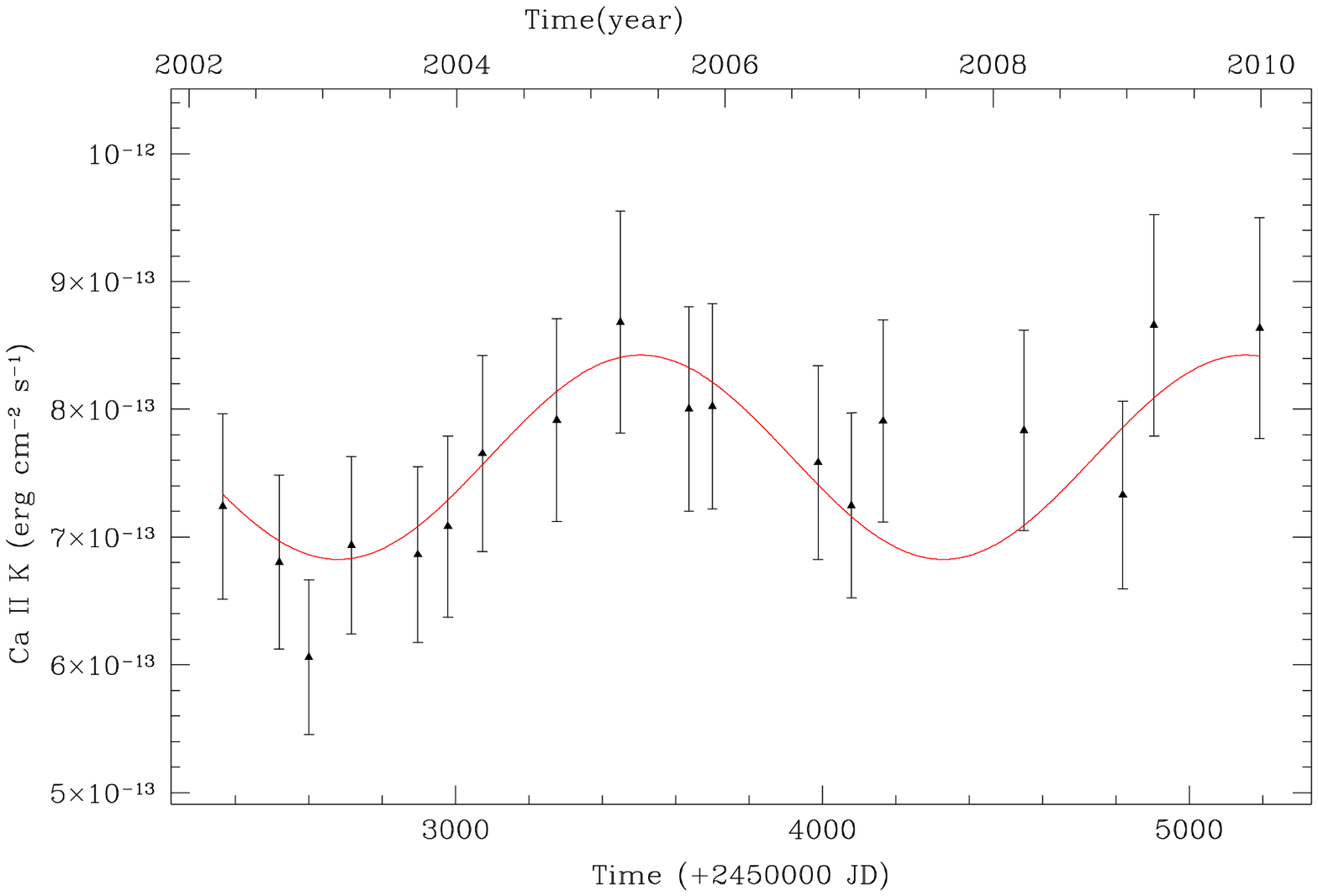}\caption{Gl 229 A. Ca \II\- K fluxes derived from the CASLEO spectra
  in Fig.~\ref{fig.sp_hd42581}.  The errors are assumed to be 10\% for
  the calibrated fluxes \citep{2004A&A...414..699C}. The solid line is
  the least-square fit to the mean annual values after establishing a
  harmonic curve of period 1649 d obtained in
  Fig. \ref{fig.hd42581_per} with a correlation coefficeint
  R=0.61.}\label{fig.hd42581_ca2_in_time}
\end{figure}

In Fig.~\ref{fig.hd42581_ca2_in_time} we plot the Ca \II\- K fluxes 
measured on the CASLEO spectra vs. time, considering an error of 10\% for the calibrated fluxes \citep{2004A&A...414..699C}. In this figure we observe
that the maximum of activity is reached in the year 2004, where the
level of activity is 43\% larger than at the minimum.

\begin{figure}[htb!]
\centering
\includegraphics[width=.7\textwidth]{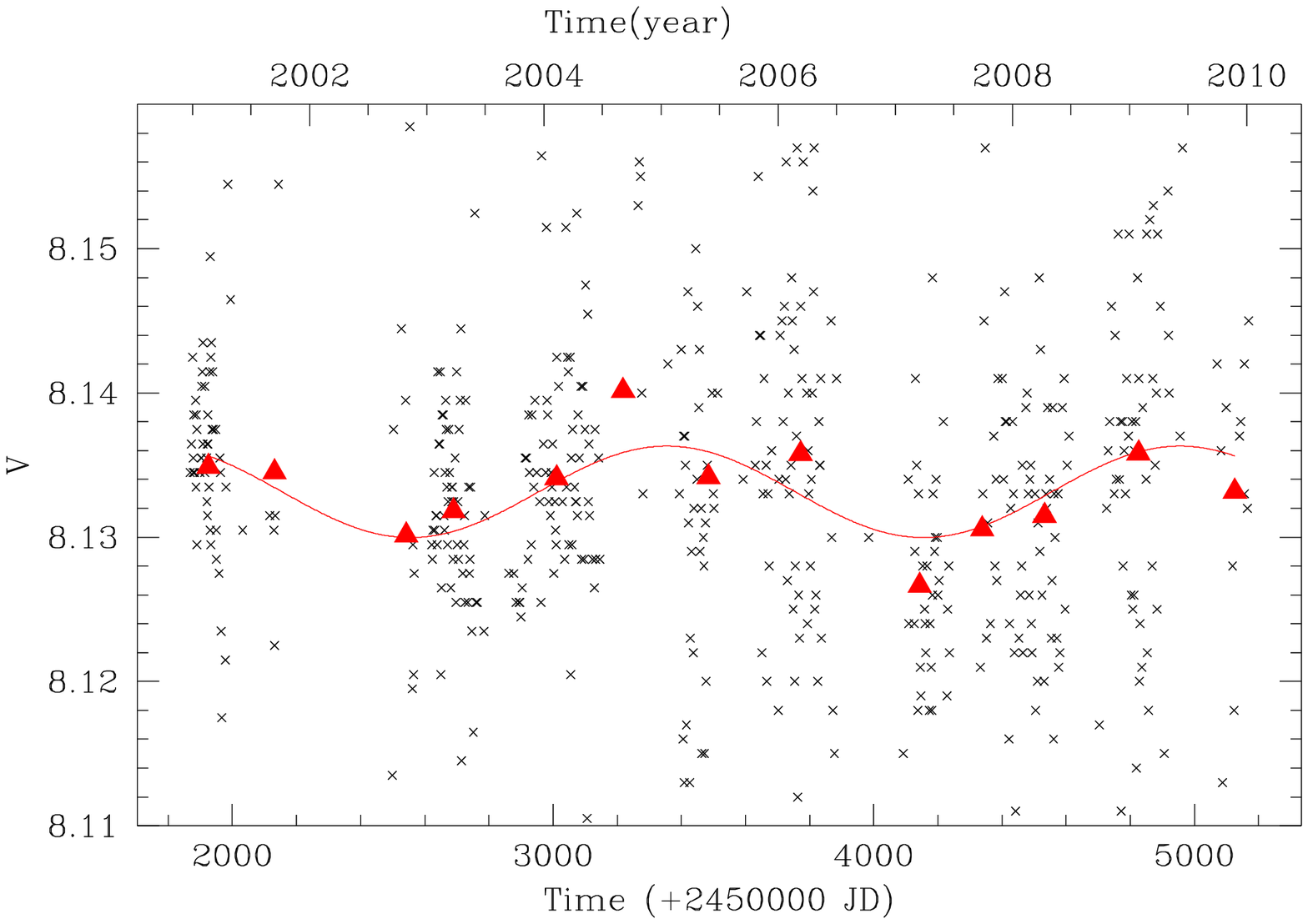}
\caption{Gl 229 A. V magnitude measured by ASAS . The solid line is the least-square fit  to the mean values ($\blacktriangle$) after establishing a  harmonic curve of period 1600 d obtained in Fig. \ref{fig.hd42581_per} with a correlation coefficeint R=0.83.}\label{fig.hd42581_asas_in_time}
\end{figure}

On the other hand, in Fig.~\ref{fig.hd42581_asas_in_time} we plot the
V magnitude 
obtained from the ASAS catalog.  The ASAS data for this system cover the
period between 2001 and 2010. We discarded all the observations that
were not qualified as either A or B in the ASAS database
(i.e. we retained only the best quality data), as well as 16 outlier
observations. The final dataset consists of 469 points,
with typical errors of around 30 mmag. The mean  magnitude
variation for the whole period is only $\sim$0.17\% around $\langle
V\rangle=8.133\pm 0.003$, while the short-scale variations in the V
magnitude have a $\sim$1\% amplitude.  To explore
if these variations are governed by spots and active regions on the
stellar surface, we analyzed if there is evidence of rotational
modulation in the ASAS data, studying the dataset in each observing
season with the Lomb-Scargle periodogram
\citep{1982ApJ...263..835S,1986ApJ...302..757H}. However, in agreement with
\cite{1985MNRAS.214..119B}, we did not find evidence of rotational
modulation in the ASAS datasets.

To search for long-term chromospheric cycles, we analyzed both
   datasets with the Lomb-Scargle periodogram, and we computed the
   False Alarm Probability (FAP) of the obtained periods with a Monte
   Carlo simulation, as is explained in \cite{2009A&A...495..287B}.
   For the Ca \II\- K fluxes plotted in
   Fig. \ref{fig.hd42581_ca2_in_time} we obtained a period
   $P_{CASLEO}=1649\pm 81$ days with a very good FAP=1.4\%. 

For the ASAS time series we obtained a 
peak at 1649 days with a FAP=$5 \times 10^{-5}$. This
extremely low FAP is related to the large number of points of this
series (see Eq. 22 in \citealt{1986ApJ...302..757H}). To reduce the
number of points, and to eliminate short-term variations, we binned
the data grouping together the observations corresponding to the same
observing season. We weighted each value with the error
reported in the ASAS database, and we computed the error of each
mean V magnitude as the square root of the variance weighted mean (see
\citealt{Frod79}, Eq. 9.12). For this series we detected a similar period,
$P_{ASAS}=1600\pm 75$ days, with a FAP=7.2\%. To check \textbf{wether} the
period we obtained depends on the binning, we also analyzed
the mean {\it annual} V magnitude with the same algorithm and we obtained
a similar peak at $P_{ASAS}=1722\pm 46$ days, altough with a larger
FAP=70\%. 

To analyze the robustness of our results, we considered the series
obtained by binning the data by observing season, and 
we analyzed the periodograms of
the series obtained by alternatively eliminating each data
point. We obtained $P_{CASLEO}$ between 1615 and 1799 days with
FAPs$<$15\% (75\% of them were $P_{CASLEO}$=1649 days) and $P_{ASAS}$
between 1493 and 1723 days with FAPs$<$45\%, and 62\% were
$P_{ASAS}=1600$ days. The fact that the periods are similar in all
cases supports the intrepretation of an harmonic component with period
$\sim$4 yr present in the ASAS series.

\begin{figure}[htb!]
\subfigure[\label{fig.hd42581_seriesp}]{\includegraphics[width=.5\textwidth]{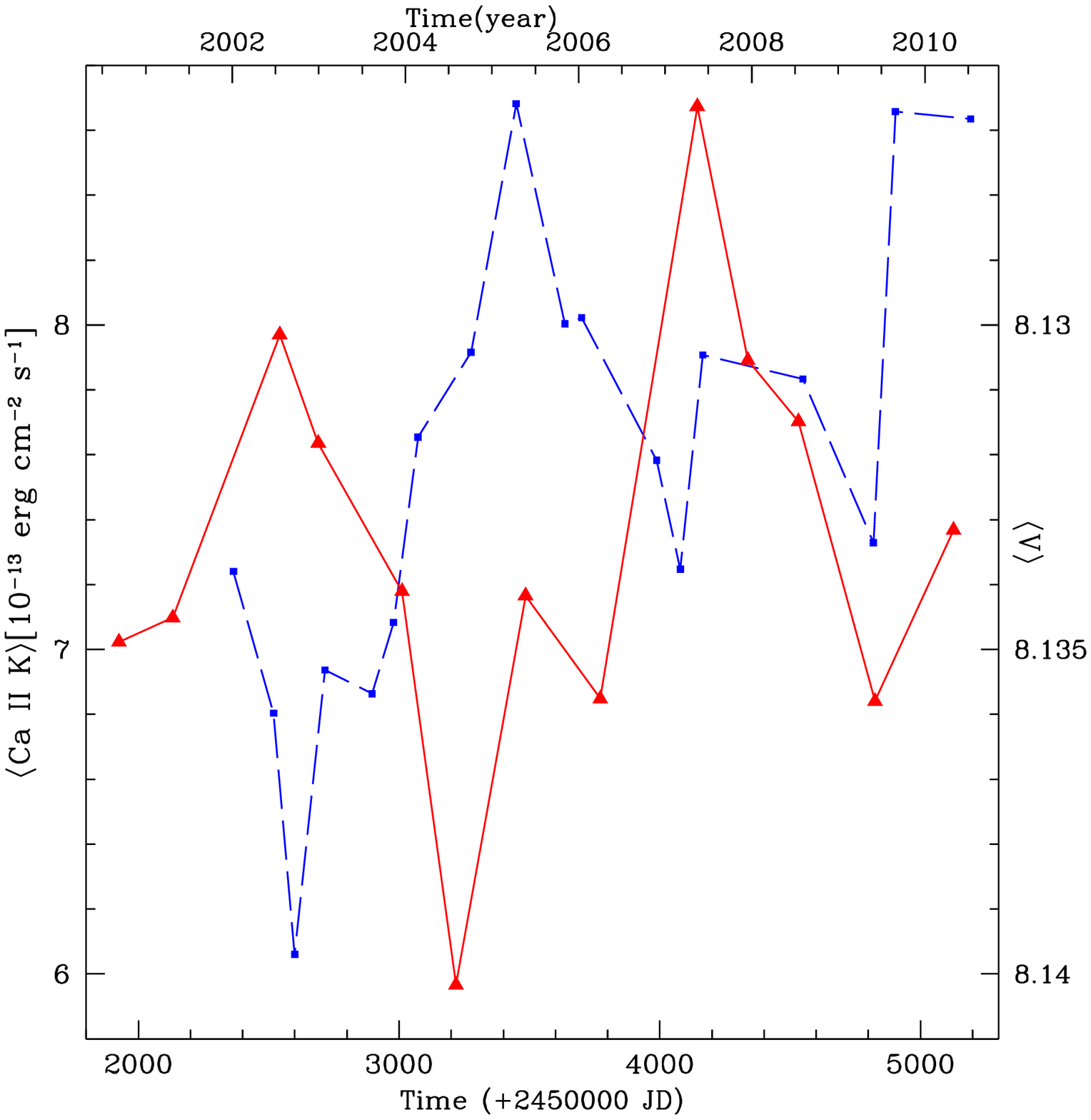}}\hfill
\subfigure[\label{fig.hd42581_per}]{\includegraphics[width=.5\textwidth]{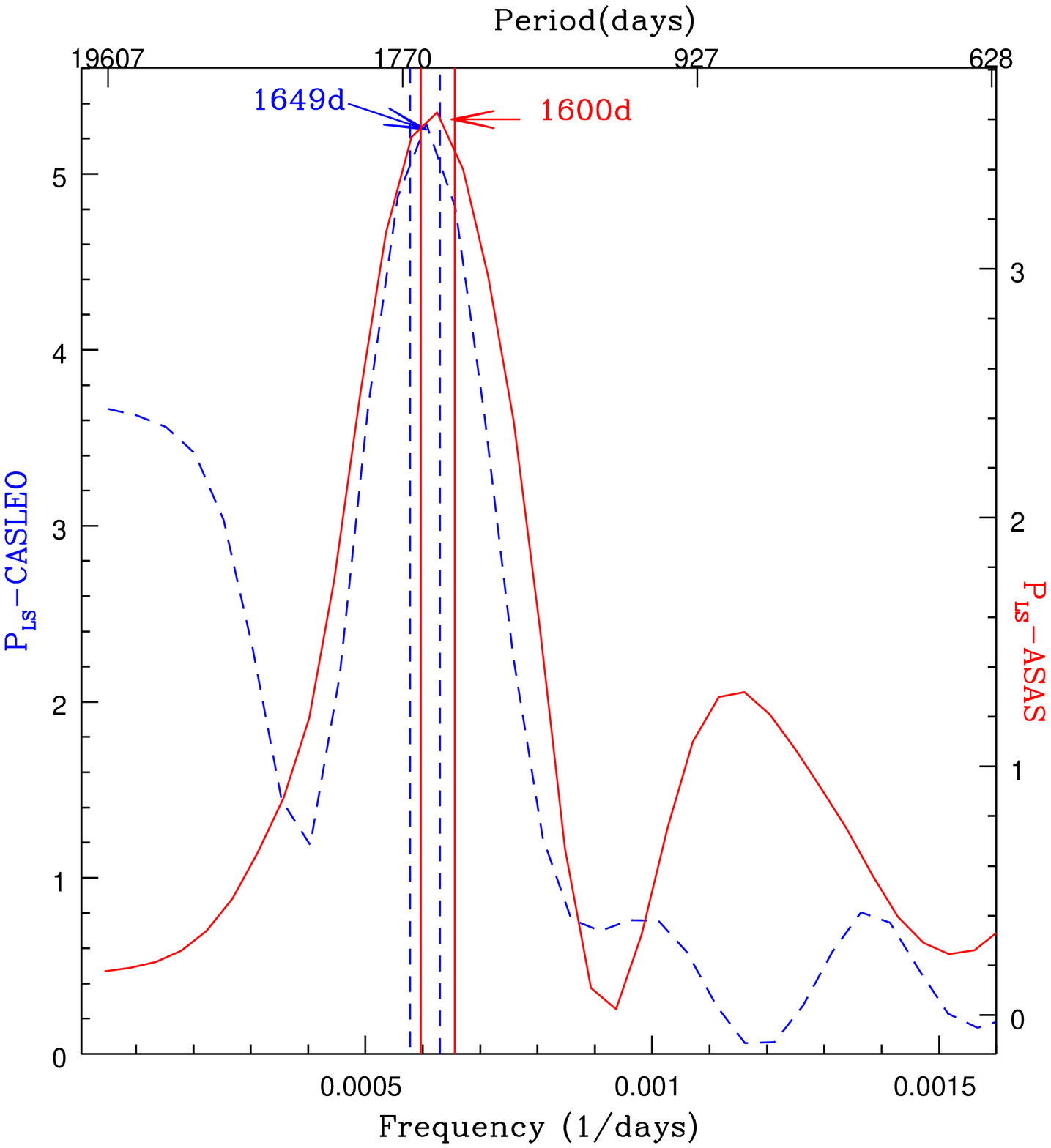}}
\caption{Gl 229 A. Left: The mean annual Ca \II\- K fluxes derived
  from Fig.~\ref{fig.hd42581_ca2_in_time} (blue, dashed) and the mean
  V magnitude \textbf{of each observing season} obtained from the data
  plotted in Fig.~\ref{fig.hd42581_asas_in_time} (red solid). To avoid
  crowding the graph, we have not included the error bars, which are
  $10\%$ for the Ca \II\- fluxes and $<0.1\%$ for the mean V
  magnitudes. Right: The \textbf{Lomb-Scargle} periodogram of the mean annual
  Ca \II\- fluxes (blue, dashed) and the mean annual V magnitude (red,
  solid). The solid and dashed vertical lines represent the error
  interval of each period.}
\end{figure}

In Fig.~\ref{fig.hd42581_seriesp} we plot the Ca \II\- fluxes and the
   weighted mean V magnitudes per observing season together. The
   Lomb-Scargle periodogram of both series are shown in
   Fig.~\ref{fig.hd42581_per}.  Note that both periods coincide within
   the statistical errors.

%%%%%\fbox{OJO que es distinto!}

\cite{1998ApJS..118..239R} studied 
long-term photometric and chromospheric emission variations in
solar-type stars, and found that young, active stars
become fainter as their chromospheric emission increases, whereas 
older, less active stars, including the Sun, tend to become brighter as their
chromospheric emission increases. 
Since during active periods the surface of the star is covered by dark spots and
bright faculae, a simple interpretation of this behavior is that long-term variability 
of young stars is spot-dominated, whereas for older stars it is
faculae-dominated. Recently, \cite{2009AJ....138..312H} 
confirmed the anti-correlation between brightness and
chromospheric emission for active stars, and found that direct correlations 
are not prevalent for the less active solar-age stars.

   In particular for Gl 229 A,
   the maximum Ca \II\- K line-core emission nearly coincides with the
   minimum mean V magnitude, as in the Sun. In
   Fig.~\ref{fig.hd42581_seriesp} we invert the V-magnitude scale and
   we observe that if we shift the chromospheric activity indicators
   by approximately 500 days, there is a good correlation
   between both series with a Pearson correlation coefficient R=0.59.

This timelag between photometric and magnetic variations has already
been observed for stars of different spectral types, from $\beta$ Com
(G0V) to $\epsilon$ Eri (K2V), including the binary $\xi$ Boo (see for
example
\citealt{1995ApJ...441..436G,1996ApJ...465..945G,1996ApJ...456..365G}).
For the coolest star in their sample ($\epsilon$ Eri) the timelag of
temperature variations is about 0.3 years ($\sim$110 days).
\cite{1996ApJ...465..945G} showed that, when different stars are
compared, this timelag is anticorrelated with effective temperature
(see their Fig. 8). However, the Sun does not fit this relation
\citep{1997ApJ...474..802G}.  Recently, \cite{2007A&A...474..345D}
reported a 140-days ($\sim$0.4 yr) timelag between chromospheric and
photospheric activity in the Gl~375 system and
\cite{2008ApJ...679.1531B} obtained a 2-year lag for the K2 III star
Arcturus. The physical explanation for these timelags remains unknown
\citep{1998ASPC..154..193G,2008ApJ...679.1531B}.

%MONTE CARLO FAP Y GLS
\section{Gl 752 A - HD 180617}\label{sec.gj752}
 Gl 752 is a binary system composed by the BY Dra variable M2.5 star
 Gl 752 A and a fainter companion Gl 752 B (also known as VB10) of
 spectral class M8V.  \cite{2009ApJ...700..623P} reported that
 the astrometric variations detected in Gl 752 B could be an evidence
 of the first giant planet around an ultra cool star. However,
 \cite{2010ApJ...711L..19B} discarded the posibility of a planetary
 companion around this star by radial velocity techniques.

%In particular, Gl 752 B probably hosts a %giant
% planet (\citealt{2009ApJ...700..623P}, but see
% \citealt{2010ApJ...711L..19B}).  In particular,
% \cite{2009ApJ...700..623P} reported that the astrometric variations
% detected in Gl 752 B could be an evidence of the first giant planet
% around an ultra cool star. However, \cite{2010ApJ...711L..19B}
% discarded the posibility of a planetary companion around this star by
% radial velocity techniques.
%

The Gl 752 system is particularly interesting to study the dynamo
processes as it is composed by two coeval stars with different
internal structures. The primary star has a radiative core, where a solar-type
dynamo could operate, and the secondary component is a
fully-convective star. With this aim, \cite{1995ApJ...455..670L} analyzed
several UV lines from \emph{HST} spectra of Gl 752 A and B. They detected a
flare-like event in the M8 component, which is an indirect evidence of
a dynamo operating in this fully-convective star. 
 
%RA 19:16:55.2569, DEC +05:10:08.054 (FORMATO APJ)

%Seg\'un SIMBAD es una flare star.
  	  
%En Cincunegui et al. (2007) se lista un $P_{rot}=16.8$d
%CARACTERISTICA DE LA ESTRELLA

\begin{figure}[htb!]
\centering
\includegraphics[width=.7\textwidth]{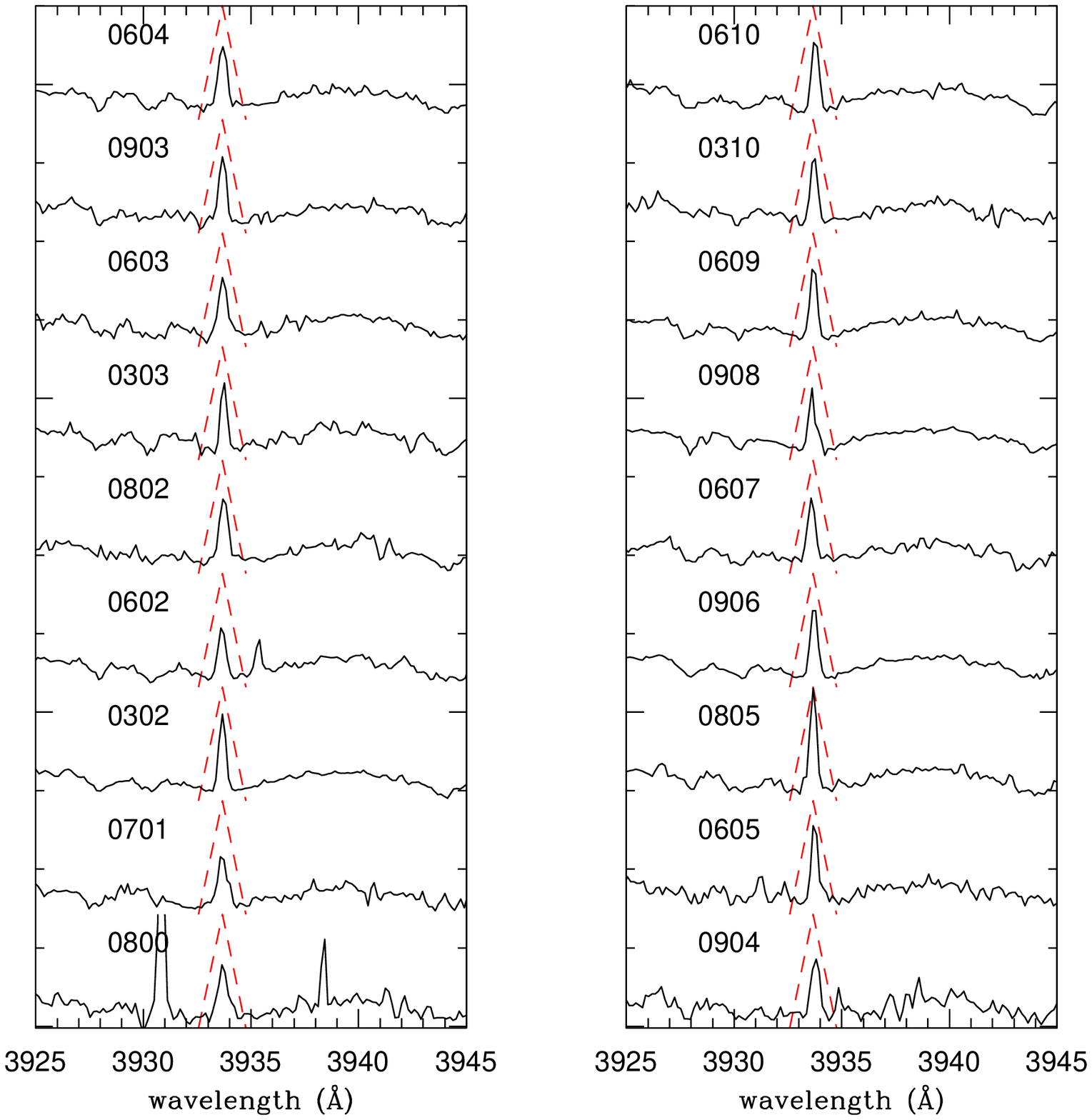}
\caption{Gl 752 A. Ca \II\- K-line in the CASLEO spectra. The triangular profiles used to integrate the fluxes are also indicated with dashed line.}\label{fig.sp_hd180617}
\end{figure}

We observed Gl 752 A at CASLEO during 18 nights between the years 2000
and 2010.  In Fig.~\ref{fig.sp_hd180617} we plot the spectra obtained,
where the labels are as in Fig.~\ref{fig.sp_hd42581} with the
month and year of the observation. The line-integration windows are
also marked with a dashed line.

%The observation logs are shown in Table \ref{tab.log_sphd180617}, and 

%Following \cite{2007astro.ph..3511C} we estimated
%$\langle logR'_{HK}\rangle=-5$ from those spectra with best
%signal-to-noise. The $R'_{HK}$-index is a measured of the
%chromospheric emission in the Ca \II\- H and K lines normalized with
%the bollometric flux, the value obtained for Gl 752 A could indicate
%that it is a poorly active star \citep{2006AJ....132..161G} which could present long-term variations
%%\citep{2007AJ....133..862H}. Nevertheless, recently
%\cite{2009AJ....138..312H} concluded that the $R'_{HK}$ is not a good
%discriminant for Maunder Minimum candidate or low-activity
%stars. Therefore, it is neccessary to study the long-term activity of Gl 752 A in
%detail to be conclusive.

\begin{figure}[htb!]
\centering
\includegraphics[width=.7\textwidth]{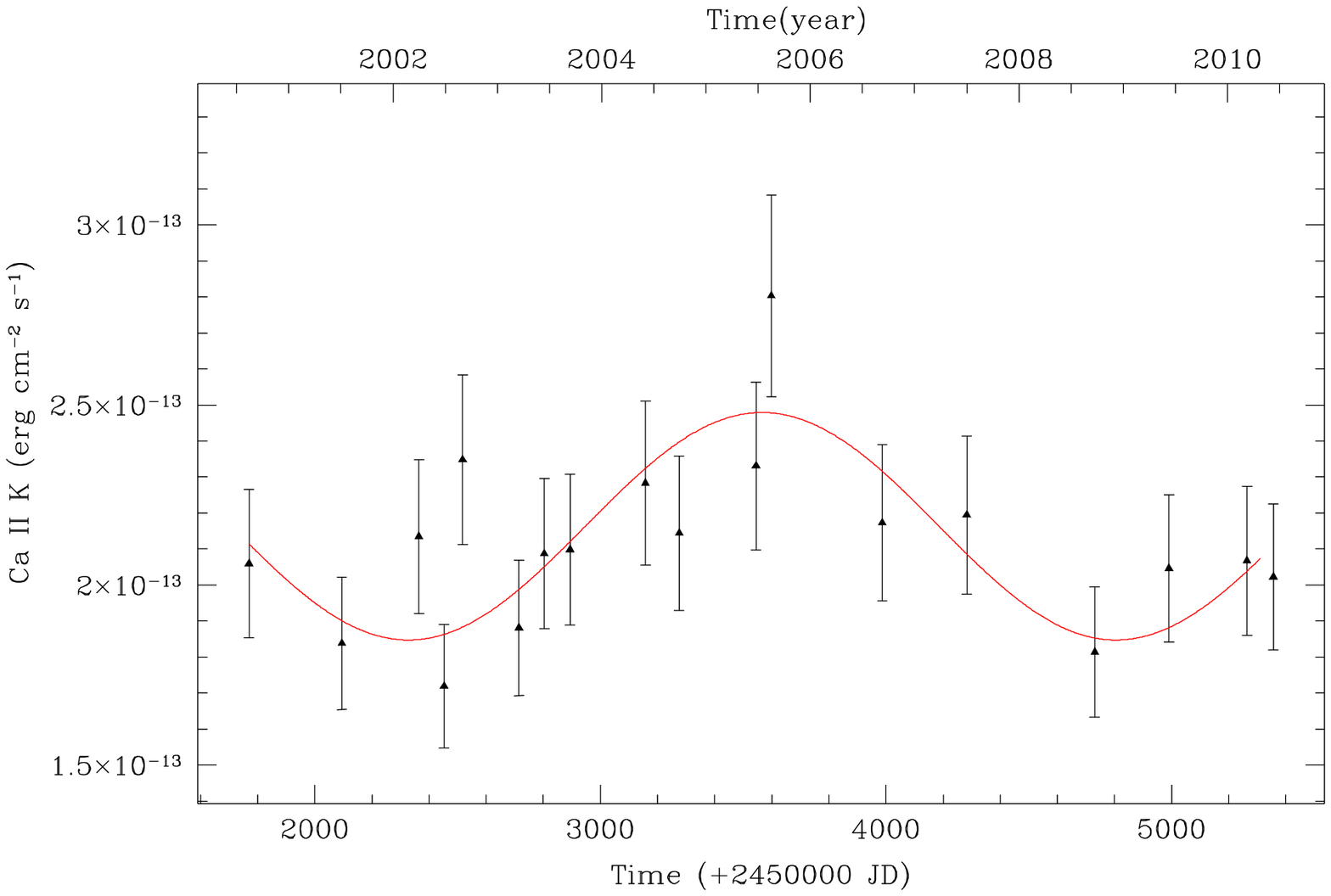}
\caption{Gl 752 A. Ca \II\- K fluxes derived from the CASLEO spectra in
  Fig.~\ref{fig.sp_hd180617}. The errors are assumed
  to be 10\% for the calibrated fluxes \citep{2004A&A...414..699C}. The solid line is the least-square fit  to the mean annual values after establishing a  harmonic curve of period 2477 d obtained in Fig. \ref{fig.hd180617_per} with a correlation coefficient R=0.79.}\label{fig.hd180617_ca2_in_time}
\end{figure}

In Fig.~\ref{fig.hd180617_ca2_in_time} we plot the Ca \II\- K fluxes
measured on the CASLEO spectra vs. time. Following
\cite{2004A&A...414..699C} we considered a 10\% error in the
calibrated fluxes. For our time interval, the yearly mean of the
K-line flux varies by 35\%, and the maximum is reached in 2005.

\begin{figure}[htb!]
\centering
\includegraphics[width=.7\textwidth]{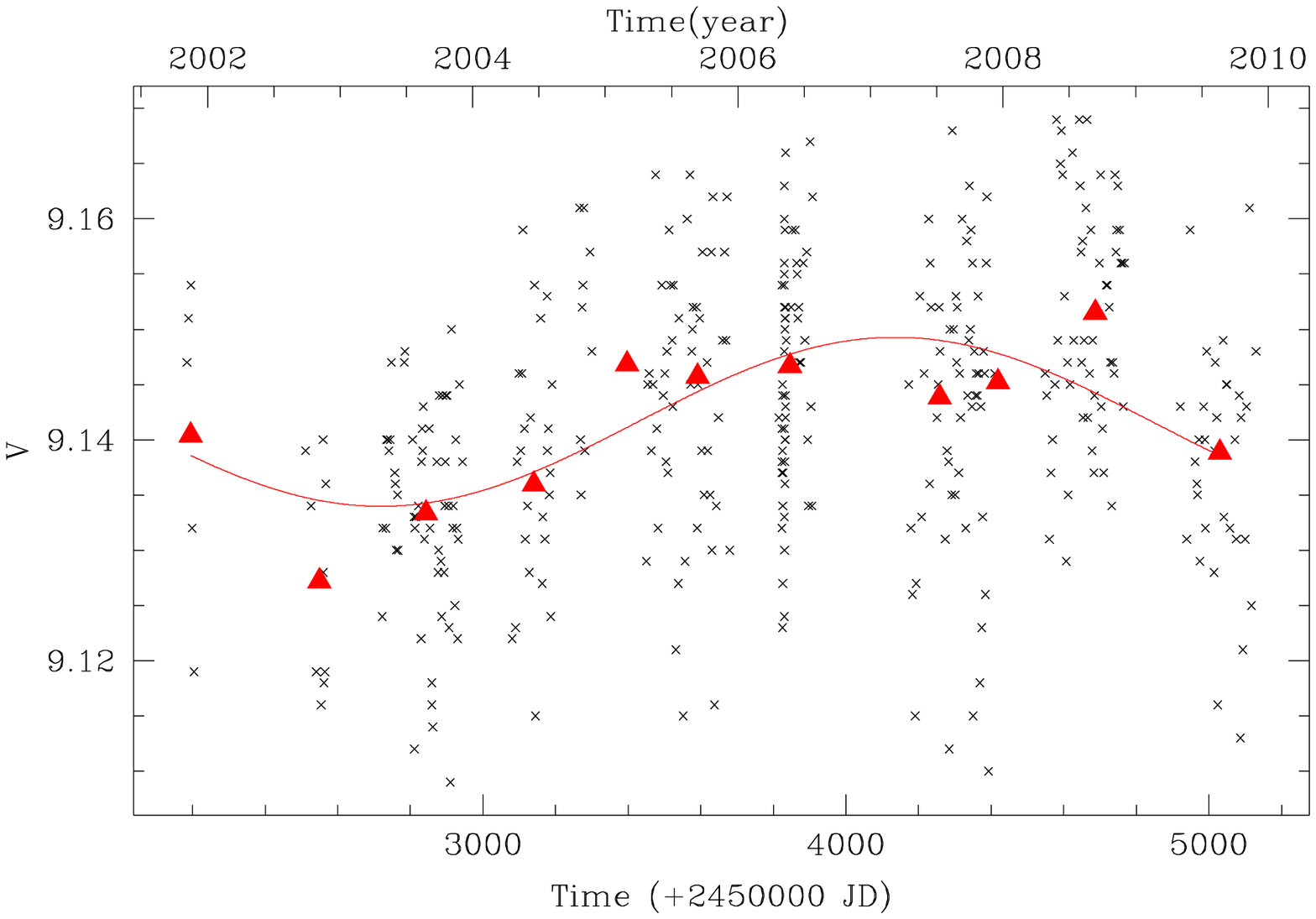}
\caption{Gl 752 A. V magnitude measured by ASAS. The solid line is the least-square fit  to the mean annual values ($\blacktriangle$) after establishing a  harmonic curve of period 2845 d obtained in Fig. \ref{fig.hd180617_per} with a corrilation coefficient R=0.77.}\label{fig.hd180617_asas_in_time}
\end{figure}

Finally, in Fig.~\ref{fig.hd180617_asas_in_time} we plot the V
magnitude available at the ASAS catalog. We selected the data of best
quality and discarded 7 outlier points, resulting in a total of 398
points. From these data, we obtain a mean magnitude $\langle
V\rangle=9.140\pm0.006$ between 2001 and 2010, with a long-term
variation of only 0.25\% between minimum and maximum. However,
short-time variations ($\sim$month) of the V magnitude are more
evident, and are between 0.3 to 1.5 \%. As done for Gl 229 A, we
analyzed if these variations were modulated by stellar rotation. In
particular, from the projected rotational velocity published in
\cite{2000AcA....50..509G}, \cite{2007astro.ph..3511C} estimated a
rotational period $P_{rot}=16.8$ days for this star. However, we did
not find any significant periodicity neither in the seasonal datasets
nor in the whole ASAS series.

To search for long-term chromospheric cycles we analyzed the Ca \II\-
K fluxes with the Lomb-Scargle periodogram, and we detected a period
$P_{CASLEO}=2510\pm 95$ days with a Monte Carlo FAP=25\%. This
relatively large FAP is due to the large variations that can be seen
in the data in 2002 and 2005. In both cases, these variations,
obtained between observations taken a few months apart, cannot be
caused by cyclic activity. If we smooth out these variations by
averaging these points, we obtain a similar period with a more
significant FAP$\sim 5\%$.

\begin{figure}[htb!]
\subfigure[\label{fig.hd180617_seriesp}]{\includegraphics[width=.5\textwidth]{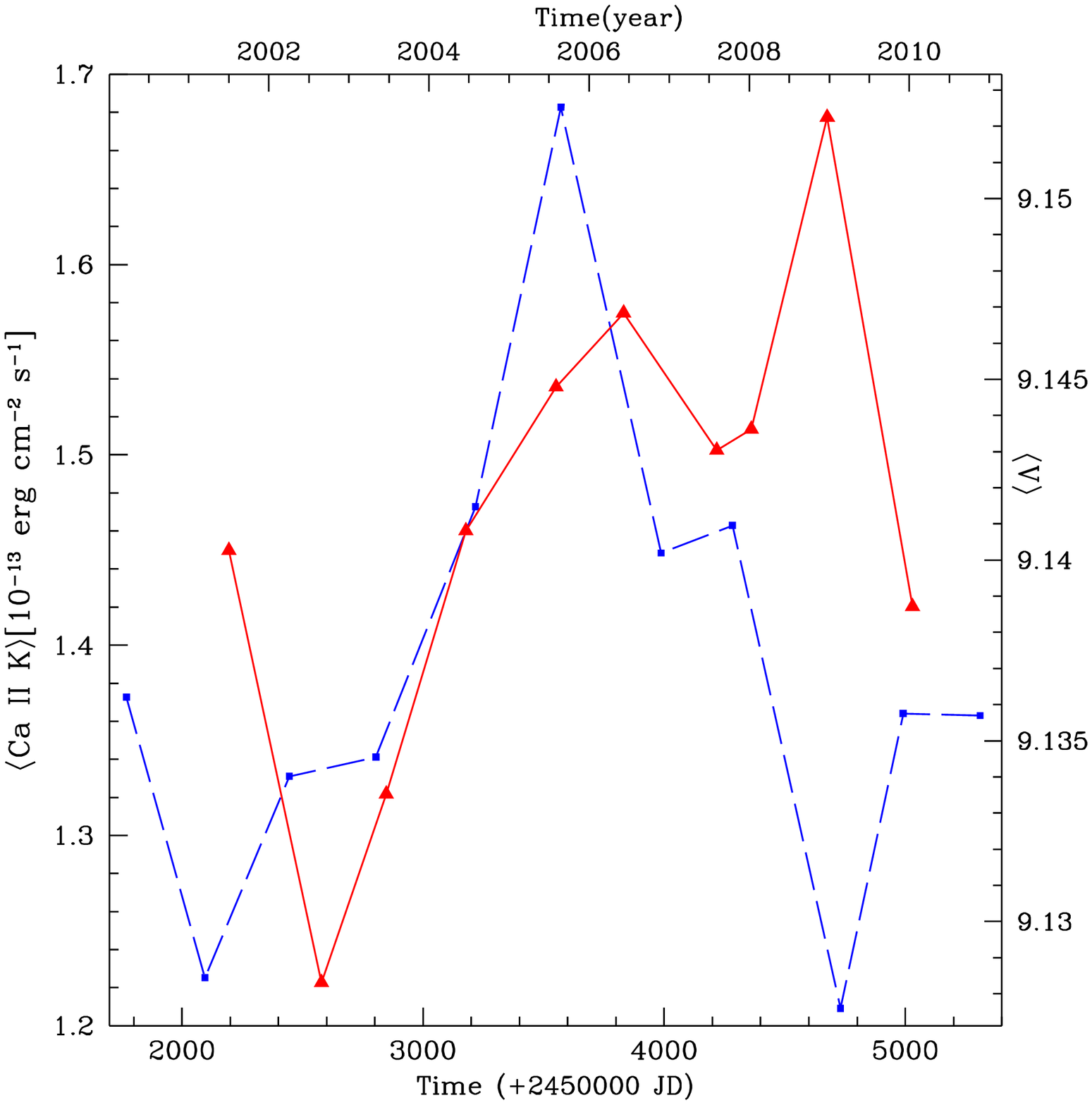}}\hfill
\subfigure[\label{fig.hd180617_per}]{\includegraphics[width=.5\textwidth]{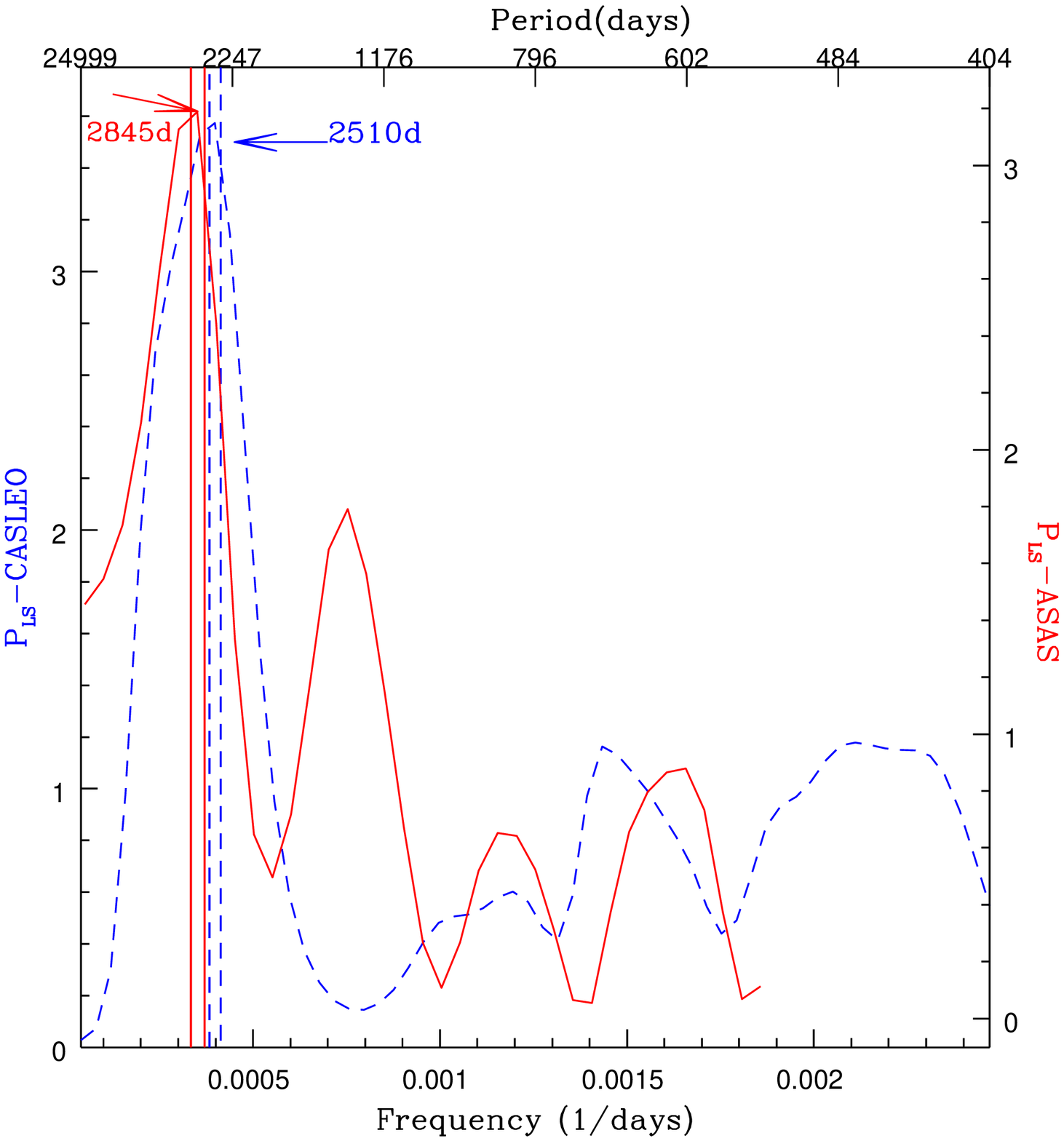}}
\caption{Gl 752 A. Left: The yearly  mean Ca \II\- fluxes derived from
  Fig.~\ref{fig.hd180617_ca2_in_time} (blue, dashed) and the weighted mean  V magnitude \textbf{of each observing season} obtained from the data
  plotted in Fig.~\ref{fig.hd180617_asas_in_time} (red, solid). To avoid crowding the graph, we have not included the error bars, which are $<10\%$ for the Ca \II\- fluxes and $<0.2\%$ for the mean V magnitudes. Right: The \textbf{Lomb-Scargle}
  periodogram of the mean  Ca \II\- fluxes (blue, dashed) and the
  mean  V magnitude (red, solid). The solid and dashed vertical lines represent the error
  interval of each period.}
\end{figure}

For the mean V magnitude per observing season the period found was
$P_{ASAS}=2845\pm 145$ days with a FAP=22\%.  In
Fig.~\ref{fig.hd180617_per} both periodograms are plotted together. In
the figure, it can be seen that both periods coincide within
2-$\sigma$ errors. The periodogram derived from the ASAS
data-sets exhibits a second peak near 1400 days, which is probably a
subharmonic component of the time-series. For completeness, we also
studied the periodogram obtained for the whole dataset, without
binning, and obtained a significant peak at 2944 days, again with an
extremely low FAP=$6\times 10^{-9}$. For the series obtained with the
annual means, we detected a significant peak at 2833 days, with a
FAP=50\%. Again, all periods coincide.
 
To study the robustness of the periods detected in
Fig.~\ref{fig.hd180617_per}, we repeated the test explained in Section
\S\ref{sec.gj229} for both series plotted in
Fig.~\ref{fig.hd180617_seriesp}. For the series obtained from the
CASLEO observations we detected $P_{CASLEO}$ between 2477 and 2505
days with FAPs$<$55\%, where nine of the eleven periods detected were
2477 days. From the ASAS dataset, we obtained eight series with
$P_{ASAS}$ between 2845 and 2990 days with FAPs$<55\%$, two series
with $P_{ASAS}$ between 3300 and 3495 days with FAPs$\sim 55\%$ and
one with $P_{ASAS}$=1398 days. We remark that in this last periodogram
there is a secondary maximum near 2985 days with FAP$\sim$50\%. This
test shows that although we shorten the series by discarding their
extremes the 7-year-period is still detected.

%\fbox{Fijate si lo que escribi aca se entiende} In order to discard
%the fact that the cyclic variations detected are due to transitory
%events as flares, we analyzed if the period obtained depends on a
%certain point or on the whole dataset. To do so, we first extract the
%first points of each time series plotted in
%Fig.~\ref{fig.hd180617_seriesp}, we analyzed the new sets with the
%Lomb-Scargle periodogram. We repeated this procedure by extracting the
%next points (one per time). Finally, for the resultant CASLEO and ASAS
%series we obtained $P_{CASLEO}$ between 2513 and 2931 days with
%FAPs<50\% and $P_{ASAS}$ between 2793 and 3258 days with FAPs<35\%
%respectively.  Therefore, the cyclic activity obtained for Gl 752 A is
%due to the long-term magnetic activity of the star.

In contrst to Gl 229 A, in Fig.~\ref{fig.hd180617_seriesp} we observe that Gl
752 A becomes fainter when the Ca \II\- emission increases, implying
that for this star spots dominate the emission.

%Since it is a moderately active star which seems to have a
%cyclic long-term variation, a spot-dominated emission is roughly
%consistent with the obtained for solar-type stars (see Fig. 8 in
%\citealt{1998ApJS..118..239R} and Fig. 9 in \citealt{2009AJ....138..312H}).}

On the other hand, in Fig.~\ref{fig.hd180617_seriesp} the Ca \II\- K
line series precedes the mean V magnitude dataset. The chromospheric
activity increased between 2001.2 and 2005.6, while the corresponding
photospheric activity increment is between 2002.7 and 2006.5, as
expressed by the Ca \II\- fluxes and the V magnitude respectively.  If
we shift the chromospheric series in 360 days, both datasets coincide
within the normalization constant with a Pearson correlation
coefficient R=0.78.

%MONTE CARLO FAP Y GLS

\section{H$\alpha$ as an activity indicator}\label{sec.halfa}
    Another activity indicator usually employed in the study of dM
stars is the flux in H$\alpha$.  The flux in this line is generally
considered to correlate well with the flux in the Ca \II\- H and K
lines and has the advantage of being located in a redder wavelength
range than the Ca \II\- lines, where M dwarfs are brighter. However,
it has been reported by \cite{2007astro.ph..3511C} that the
correlation between Ca \II\- and H$\alpha$ is not always valid. \cite{2007astro.ph..3511C} found that while some stars exhibit correlations
between H$\alpha$ and the Ca \II\- lines, the slopes change from star
to star.  Furthermore, in several cases both fluxes were not
correlated, as in the binary system Gl 375
\citep{2007A&A...474..345D}, and other stars even exhibit
anti-correlations.  Recently, \cite{2009AJ....137.3297W} analyzed the
relation between simultaneous measurements of Ca \II\- K and H$\alpha$
fluxes in a sample of nearby M3 dwarfs, and found that the
relationship between both tracers remains ambiguous for weak and
intermediate activity stars.  On the other hand,
\cite{2009A&A...501.1103M} studied the contribution of plages and
filaments in the H$\alpha$-Ca \II\- relation in the Sun for different
time scales, in an attempt to explain the results by
\cite{2007astro.ph..3511C}.

\begin{figure}[htb!]
\centering
\includegraphics[width=0.7\textwidth]{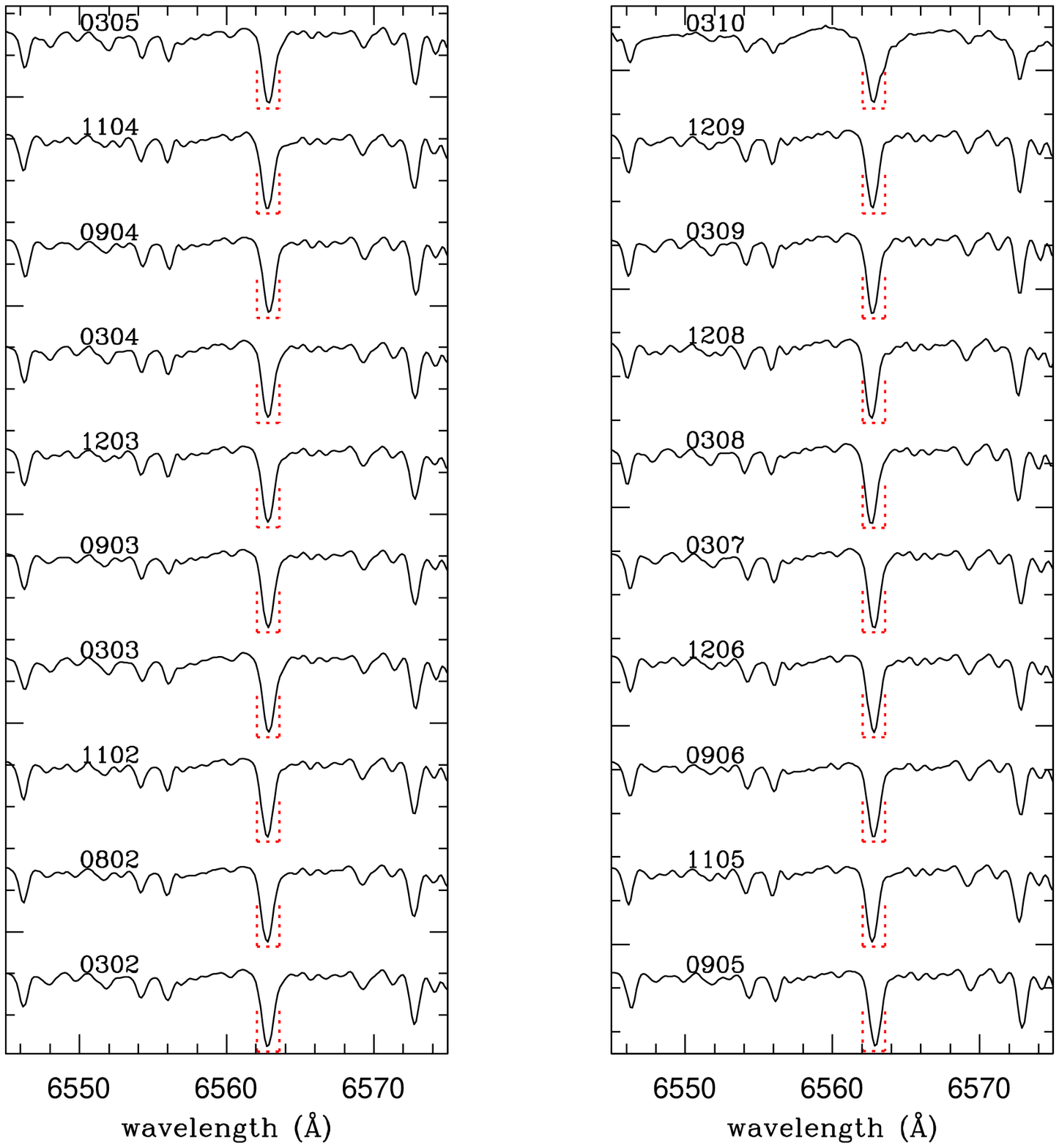}
\caption{H$\alpha$ line in the CASLEO spectra of Gl 229 A. The square integration windows are also indicated with dashed lines. }\label{fig.hd42581_ha}
\end{figure}

\begin{figure}[htb!]
\centering
{\includegraphics[width=0.7\textwidth]{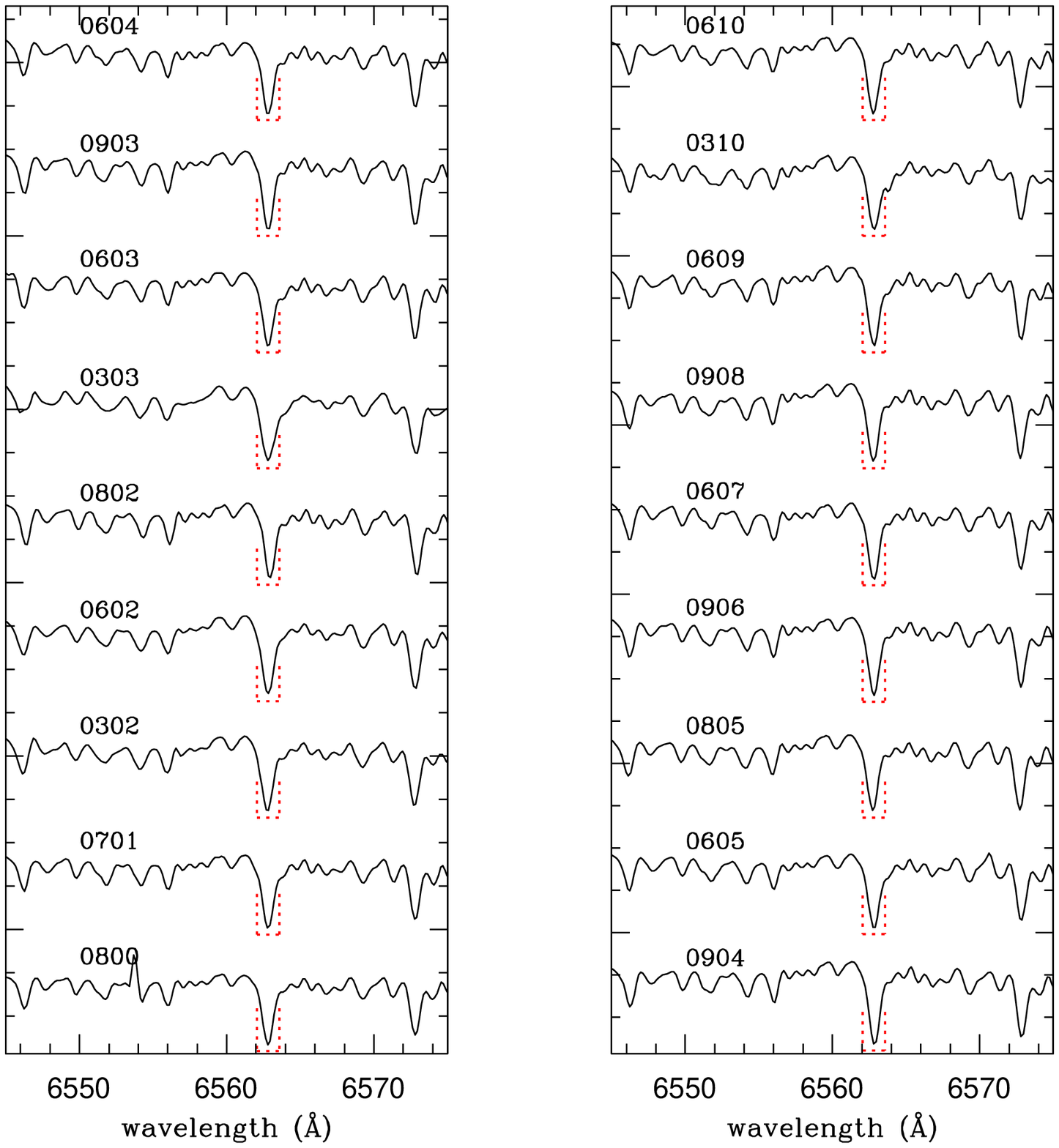}}
\caption{H$\alpha$ line in the CASLEO spectra of Gl 752 A. The square integration windows are also indicated with dashed lines. }\label{fig.hd180617_ha}
\end{figure}

\begin{figure}[htb!]
\subfigure[\label{fig.hd42581_haca}]{\includegraphics[width=.5\textwidth]{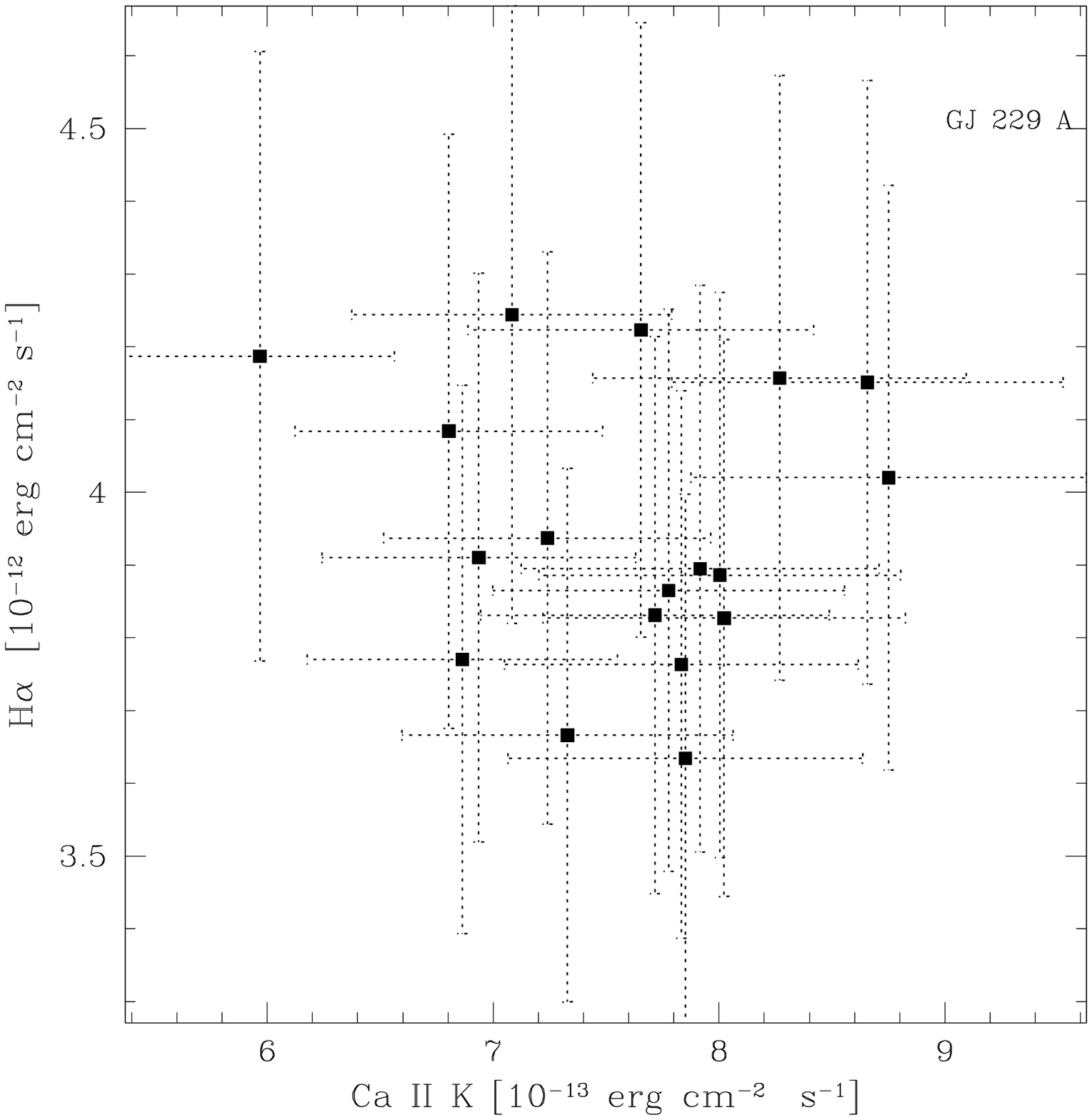}}
\subfigure[\label{fig.hd180617_haca}]{\includegraphics[width=.5\textwidth]{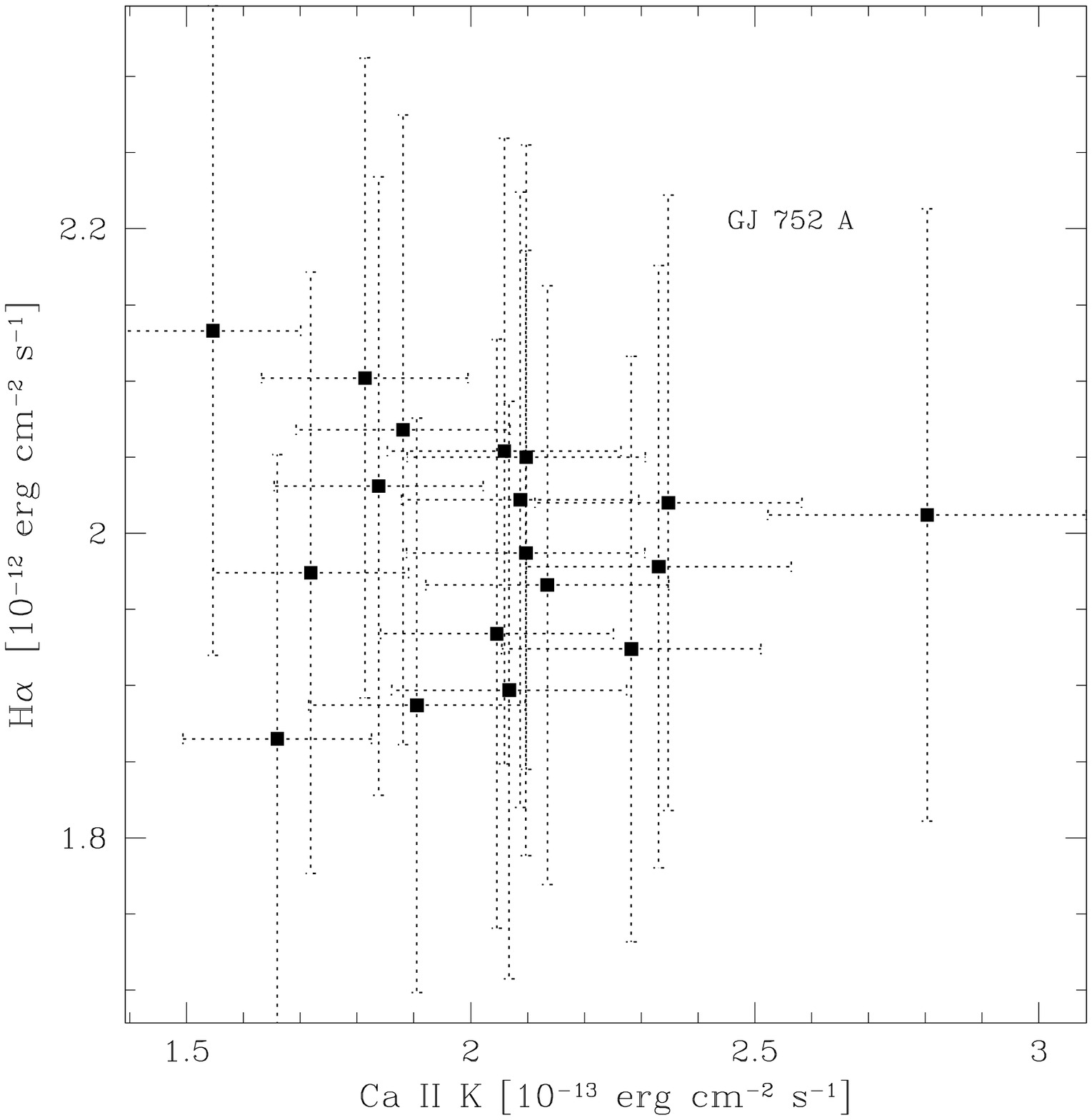}}\hfill
\caption{H$\alpha$ vs. Ca \II.   We plot the H$\alpha$
  line-core flux integrated in a 1.5 \AA\- square passband centered in
  the line vs. the Ca \II\- K flux obtained with a triangular profile
  of 1.09\AA\- FWHM for the stars Gl 229 A (left) and Gl 752 A (right). The Pearson correlation coefficients between the fluxes are
  R=-0.09 and R=-0.11 respectively. }
\end{figure}

In this work, we checked whether an H$\alpha$-Ca \II\- correlation
exists for Gl 229 A and Gl 752 A. To do so, we computed the flux in
the H$\alpha$ line as the average surface flux in a 1.5 \AA\- square
passband centered in 6562.8 \AA\- for all the spectra plotted in
Fig. \ref{fig.hd42581_ha} and Fig. \ref{fig.hd180617_ha}. \textbf{For
both stars the H$\alpha$ line is in absorption, which indicate that Gl
229 A and Gl 752 A are not strongly active, since most active M dwarfs
have H$\alpha$ in emission.} In Fig.~\ref{fig.hd180617_haca} and
\ref{fig.hd42581_haca} we plot the line-core fluxes in H$\alpha$
vs. those in the Ca \II\- K line. It can be seen that, within the
errors, none of these stars exhibit any sign of correlation between
the lines fluxes, as expressed by the Pearson correlation coefficients
R=-0.09 for Gl 229 A and R=-0.11 for Gl 752 A. Therefore, the
H$\alpha$ line cannot be used as an activity indicator for these
stars.

\section{Conclusions}
In this work we studied the long-term activity of two M dwarf stars:
Gl 229 A and Gl 752 A. To do so, we analyzed the Ca \II\- K
line-core fluxes measured on our spectra obtained at the
CASLEO Observatory since 2000, and the ASAS photometric data.
Using the Lomb-Scargle periodogram, we obtained a possible activity cycle of
$\sim$4 and $\sim$7 yrs for Gl 229 A and Gl 752 A, respectively.
We note that, for both stars, similar periods are found  in
the photometry and in the Ca \II\- fluxes, which constitute two completely
independent datasets, a fact which strongly reinforces the significance
of the results. As our program continues, we hope to confirm
both periods in future work.

These periods are in agreement with evidences of periodic activity
 in M dwarfs which was previously reported for the dM stars
Proxima Centauri \citep{2007A&A...461.1107C}
and the Gl 375 system \citep{2007A&A...474..345D}.

Another interesting result of this work is that the chromospheric
activity seems to precede the photometric data in $\sim1$ yr.  This
time-lag was also reported in the dMe system Gl 375
\citep{2007A&A...474..345D}. Although the physical explanation for
this phenomenon remains unknown, one possible interpretation for the
offset between the photometric and chromospheric variations is that
the chromosphere gets heated before spots are formed. The time-lag
detected in Gl 229 A and Gl 752 A's data-sets bring further evidences
of this phenomenon in late-type stars.

On the other hand, since the chromospheric emission in Gl 752 A is
larger when the system is fainter, we conclude that the emission of
this star should be dominated by dark spots rather than bright active
regions. In contrast, in the case of Gl 229 A, the data evidence that
the active regions dominate the emission. These results are consistent
with the chromospheric-photospheric relations observed in solar-type
stars in the literature
\citep{1998ApJS..118..239R,2007ApJS..171..260L,2009AJ....138..312H}.

We also analyzed simultaneous measurements of Ca \II\- K and H$\alpha$
fluxes obtained from our CASLEO spectra, and found no evident
correlation between both indices, in agreement with previous results
by \cite{2007astro.ph..3511C}. Also, \cite{2009AJ....137.3297W} found
some correlation for the most active stars, but it breaks down for
weakly active stars like the ones we study in the present paper.

\acknowledgments The CCD and data acquisition system at CASLEO has
 been partly financed by R. M. Rich through U.S. NSF grant
 AST-90-15827. This research has made use of the SIMBAD database,
 operated at CDS, Strasbourg, France. We would like to thank the
 CASLEO staff and Mr. Pablo Valenzuela, for his invaluable help
 in the data reduction.

\bibliographystyle{apj}
%\bibliography{bibliografia2}

\begin{thebibliography}{49}
\expandafter\ifx\csname natexlab\endcsname\relax\def\natexlab#1{#1}\fi

\bibitem[{{Baliunas} {et~al.}(1995){Baliunas}, {Donahue}, {Soon}, {Horne},
  {Frazer}, {Woodard-Eklund}, {Bradford}, {Rao}, {Wilson}, {Zhang}, {Bennett},
  {Briggs}, {Carroll}, {Duncan}, {Figueroa}, {Lanning}, {Misch}, {Mueller},
  {Noyes}, {Poppe}, {Porter}, {Robinson}, {Russell}, {Shelton}, {Soyumer},
  {Vaughan}, \& {Whitney}}]{1995ApJ...438..269B}
{Baliunas}, S.~L., {Donahue}, R.~A., {Soon}, W.~H., {Horne}, J.~H., {Frazer},
  J., {Woodard-Eklund}, L., {Bradford}, M., {Rao}, L.~M., {Wilson}, O.~C.,
  {Zhang}, Q., {Bennett}, W., {Briggs}, J., {Carroll}, S.~M., {Duncan}, D.~K.,
  {Figueroa}, D., {Lanning}, H.~H., {Misch}, T., {Mueller}, J., {Noyes}, R.~W.,
  {Poppe}, D., {Porter}, A.~C., {Robinson}, C.~R., {Russell}, J., {Shelton},
  J.~C., {Soyumer}, T., {Vaughan}, A.~H., \& {Whitney}, J.~H. 1995, \apj, 438,
  269

\bibitem[{{Barnes}(2003)}]{2003ApJ...586..464B}
{Barnes}, S.~A. 2003, \apj, 586, 464

\bibitem[{{Bean} {et~al.}(2010){Bean}, {Seifahrt}, {Hartman}, {Nilsson},
  {Reiners}, {Dreizler}, {Henry}, \& {Wiedemann}}]{2010ApJ...711L..19B}
{Bean}, J.~L., {Seifahrt}, A., {Hartman}, H., {Nilsson}, H., {Reiners}, A.,
  {Dreizler}, S., {Henry}, T.~J., \& {Wiedemann}, G. 2010, \apjl, 711, L19

\bibitem[{{Bonfils} {et~al.}(2007){Bonfils}, {Mayor}, {Delfosse}, {Forveille},
  {Gillon}, {Perrier}, {Udry}, {Bouchy}, {Lovis}, {Pepe}, {Queloz}, {Santos},
  \& {Bertaux}}]{2007A&A...474..293B}
{Bonfils}, X., {Mayor}, M., {Delfosse}, X., {Forveille}, T., {Gillon}, M.,
  {Perrier}, C., {Udry}, S., {Bouchy}, F., {Lovis}, C., {Pepe}, F., {Queloz},
  D., {Santos}, N.~C., \& {Bertaux}, J. 2007, \aap, 474, 293

\bibitem[{{Brown} {et~al.}(2008){Brown}, {Gray}, \&
  {Baliunas}}]{2008ApJ...679.1531B}
{Brown}, K.~I.~T., {Gray}, D.~F., \& {Baliunas}, S.~L. 2008, \apj, 679, 1531

\bibitem[{{Browning}(2008)}]{2008ApJ...676.1262B}
{Browning}, M.~K. 2008, \apj, 676, 1262

\bibitem[{{Browning} {et~al.}(2010){Browning}, {Basri}, {Marcy}, {West}, \&
  {Zhang}}]{2010AJ....139..504B}
{Browning}, M.~K., {Basri}, G., {Marcy}, G.~W., {West}, A.~A., \& {Zhang}, J.
  2010, \aj, 139, 504

\bibitem[{{Buccino} {et~al.}(2006){Buccino}, {Lemarchand}, \&
  {Mauas}}]{2006Icar..183..491B}
{Buccino}, A.~P., {Lemarchand}, G.~A., \& {Mauas}, P.~J.~D. 2006, Icarus, 183,
  491

\bibitem[{{Buccino} {et~al.}(2007){Buccino}, {Lemarchand}, \&
  {Mauas}}]{2007Icar..192..582B}
---. 2007, Icarus, 192, 582

\bibitem[{{Buccino} \& {Mauas}(2008)}]{2008A&A...483..903B}
{Buccino}, A.~P., \& {Mauas}, P.~J.~D. 2008, \aap, 483, 903

\bibitem[{{Buccino} \& {Mauas}(2009)}]{2009A&A...495..287B}
---. 2009, \aap, 495, 287

\bibitem[{{Byrne} {et~al.}(1985){Byrne}, {Doyle}, \&
  {Menzies}}]{1985MNRAS.214..119B}
{Byrne}, P.~B., {Doyle}, J.~G., \& {Menzies}, J.~W. 1985, \mnras, 214, 119

\bibitem[{{Chabrier} \& {K{\"u}ker}(2006)}]{2006A&A...446.1027C}
{Chabrier}, G., \& {K{\"u}ker}, M. 2006, \aap, 446, 1027

\bibitem[{{Charbonneau} {et~al.}(2009){Charbonneau}, {Berta}, {Irwin}, {Burke},
  {Nutzman}, {Buchhave}, {Lovis}, {Bonfils}, {Latham}, {Udry}, {Murray-Clay},
  {Holman}, {Falco}, {Winn}, {Queloz}, {Pepe}, {Mayor}, {Delfosse}, \&
  {Forveille}}]{2009Natur.462..891C}
{Charbonneau}, D., {Berta}, Z.~K., {Irwin}, J., {Burke}, C.~J., {Nutzman}, P.,
  {Buchhave}, L.~A., {Lovis}, C., {Bonfils}, X., {Latham}, D.~W., {Udry}, S.,
  {Murray-Clay}, R.~A., {Holman}, M.~J., {Falco}, E.~E., {Winn}, J.~N.,
  {Queloz}, D., {Pepe}, F., {Mayor}, M., {Delfosse}, X., \& {Forveille}, T.
  2009, \nat, 462, 891

\bibitem[{{Cincunegui} {et~al.}(2007{\natexlab{a}}){Cincunegui}, {D{\'{\i}}az},
  \& {Mauas}}]{2007A&A...461.1107C}
{Cincunegui}, C., {D{\'{\i}}az}, R.~F., \& {Mauas}, P.~J.~D.
  2007{\natexlab{a}}, \aap, 461, 1107

\bibitem[{{Cincunegui} {et~al.}(2007{\natexlab{b}}){Cincunegui}, {D{\'{\i}}az},
  \& {Mauas}}]{2007astro.ph..3511C}
---. 2007{\natexlab{b}}, \aap, 469, 309

\bibitem[{{Cincunegui} \& {Mauas}(2004)}]{2004A&A...414..699C}
{Cincunegui}, C., \& {Mauas}, P.~J.~D. 2004, \aap, 414, 699

\bibitem[{{Delfosse} {et~al.}(1998){Delfosse}, {Forveille}, {Perrier}, \&
  {Mayor}}]{1998A&A...331..581D}
{Delfosse}, X., {Forveille}, T., {Perrier}, C., \& {Mayor}, M. 1998, \aap, 331,
  581

\bibitem[{{D{\'{\i}}az} {et~al.}(2007{\natexlab{a}}){D{\'{\i}}az},
  {Cincunegui}, \& {Mauas}}]{2007MNRAS.378.1007D}
{D{\'{\i}}az}, R.~F., {Cincunegui}, C., \& {Mauas}, P.~J.~D.
  2007{\natexlab{a}}, \mnras, 378, 1007

\bibitem[{{D{\'{\i}}az} {et~al.}(2007{\natexlab{b}}){D{\'{\i}}az},
  {Gonz{\'a}lez}, {Cincunegui}, \& {Mauas}}]{2007A&A...474..345D}
{D{\'{\i}}az}, R.~F., {Gonz{\'a}lez}, J.~F., {Cincunegui}, C., \& {Mauas},
  P.~J.~D. 2007{\natexlab{b}}, \aap, 474, 345

\bibitem[{{Frodesen} {et~al.}(1979){Frodesen}, {Skjeggestad}, \&
  {Tofte}}]{Frod79}
{Frodesen}, G.~A., {Skjeggestad}, O., \& {Tofte}, H. 1979, {Probability and
  Statistics in Particle Physics} (Universitetsforlaget)

\bibitem[{{Glebocki} \& {Stawikowski}(2000)}]{2000AcA....50..509G}
{Glebocki}, R., \& {Stawikowski}, A. 2000, Acta Astronomica, 50, 509

\bibitem[{{Gray}(1998)}]{1998ASPC..154..193G}
{Gray}, D.~F. 1998, in Astronomical Society of the Pacific Conference Series,
  Vol. 154, Cool Stars, Stellar Systems, and the Sun, ed. {R.~A.~Donahue \&
  J.~A.~Bookbinder}, 193--+

\bibitem[{{Gray} \& {Baliunas}(1995)}]{1995ApJ...441..436G}
{Gray}, D.~F., \& {Baliunas}, S.~L. 1995, \apj, 441, 436

\bibitem[{{Gray} {et~al.}(1996{\natexlab{a}}){Gray}, {Baliunas}, {Lockwood}, \&
  {Skiff}}]{1996ApJ...465..945G}
{Gray}, D.~F., {Baliunas}, S.~L., {Lockwood}, G.~W., \& {Skiff}, B.~A.
  1996{\natexlab{a}}, \apj, 465, 945

\bibitem[{{Gray} {et~al.}(1996{\natexlab{b}}){Gray}, {Baliunas}, {Lockwood}, \&
  {Skiff}}]{1996ApJ...456..365G}
---. 1996{\natexlab{b}}, \apj, 456, 365

\bibitem[{{Gray} \& {Livingston}(1997)}]{1997ApJ...474..802G}
{Gray}, D.~F., \& {Livingston}, W.~C. 1997, \apj, 474, 802

\bibitem[{{Hall} {et~al.}(2009){Hall}, {Henry}, {Lockwood}, {Skiff}, \&
  {Saar}}]{2009AJ....138..312H}
{Hall}, J.~C., {Henry}, G.~W., {Lockwood}, G.~W., {Skiff}, B.~A., \& {Saar},
  S.~H. 2009, \aj, 138, 312

\bibitem[{{Hawley} \& {Pettersen}(1991)}]{1991ApJ...378..725H}
{Hawley}, S.~L., \& {Pettersen}, B.~R. 1991, \apj, 378, 725

\bibitem[{{Horne} \& {Baliunas}(1986)}]{1986ApJ...302..757H}
{Horne}, J.~H., \& {Baliunas}, S.~L. 1986, \apj, 302, 757

\bibitem[{{Johns-Krull} \& {Valenti}(1996)}]{1996ApJ...459L..95J}
{Johns-Krull}, C.~M., \& {Valenti}, J.~A. 1996, \apjl, 459, L95+

\bibitem[{{Linsky} {et~al.}(1995){Linsky}, {Wood}, {Brown}, {Giampapa}, \&
  {Ambruster}}]{1995ApJ...455..670L}
{Linsky}, J.~L., {Wood}, B.~E., {Brown}, A., {Giampapa}, M.~S., \& {Ambruster},
  C. 1995, \apj, 455, 670

\bibitem[{{Lockwood} {et~al.}(2007){Lockwood}, {Skiff}, {Henry}, {Henry},
  {Radick}, {Baliunas}, {Donahue}, \& {Soon}}]{2007ApJS..171..260L}
{Lockwood}, G.~W., {Skiff}, B.~A., {Henry}, G.~W., {Henry}, S., {Radick},
  R.~R., {Baliunas}, S.~L., {Donahue}, R.~A., \& {Soon}, W. 2007, \apjs, 171,
  260

\bibitem[{{Mauas} \& {Falchi}(1994)}]{1994A&A...281..129M}
{Mauas}, P.~J.~D., \& {Falchi}, A. 1994, \aap, 281, 129

\bibitem[{{Mauas} \& {Falchi}(1996)}]{1996A&A...310..245M}
---. 1996, \aap, 310, 245

\bibitem[{{Mauas} {et~al.}(1997){Mauas}, {Falchi}, {Pasquini}, \&
  {Pallavicini}}]{1997A&A...326..249M}
{Mauas}, P.~J.~D., {Falchi}, A., {Pasquini}, L., \& {Pallavicini}, R. 1997,
  \aap, 326, 249

\bibitem[{{Meunier} \& {Delfosse}(2009)}]{2009A&A...501.1103M}
{Meunier}, N., \& {Delfosse}, X. 2009, \aap, 501, 1103

\bibitem[{{Mochnacki} \& {Zirin}(1980)}]{1980ApJ...239L..27M}
{Mochnacki}, S.~W., \& {Zirin}, H. 1980, \apjl, 239, L27

\bibitem[{{Nakajima} {et~al.}(1995){Nakajima}, {Oppenheimer}, {Kulkarni},
  {Golimowski}, {Matthews}, \& {Durrance}}]{1995Natur.378..463N}
{Nakajima}, T., {Oppenheimer}, B.~R., {Kulkarni}, S.~R., {Golimowski}, D.~A.,
  {Matthews}, K., \& {Durrance}, S.~T. 1995, \nat, 378, 463

\bibitem[{{Pace} \& {Pasquini}(2004)}]{2004A&A...426.1021P}
{Pace}, G., \& {Pasquini}, L. 2004, \aap, 426, 1021

\bibitem[{{Parker}(1955)}]{1955ApJ...122..293P}
{Parker}, E.~N. 1955, \apj, 122, 293

\bibitem[{{Pojmanski}(2002)}]{2002AcA....52..397P}
{Pojmanski}, G. 2002, Acta Astronomica, 52, 397

\bibitem[{{Pravdo} \& {Shaklan}(2009)}]{2009ApJ...700..623P}
{Pravdo}, S.~H., \& {Shaklan}, S.~B. 2009, \apj, 700, 623

\bibitem[{{Radick} {et~al.}(1998){Radick}, {Lockwood}, {Skiff}, \&
  {Baliunas}}]{1998ApJS..118..239R}
{Radick}, R.~R., {Lockwood}, G.~W., {Skiff}, B.~A., \& {Baliunas}, S.~L. 1998,
  \apjs, 118, 239

\bibitem[{{Reiners} \& {Basri}(2008)}]{2008ApJ...684.1390R}
{Reiners}, A., \& {Basri}, G. 2008, \apj, 684, 1390

\bibitem[{{Scargle}(1982)}]{1982ApJ...263..835S}
{Scargle}, J.~D. 1982, \apj, 263, 835

\bibitem[{{Vaughan} {et~al.}(1978){Vaughan}, {Preston}, \&
  {Wilson}}]{1978PASP...90..267V}
{Vaughan}, A.~H., {Preston}, G.~W., \& {Wilson}, O.~C. 1978, \pasp, 90, 267

\bibitem[{{Walkowicz} \& {Hawley}(2009)}]{2009AJ....137.3297W}
{Walkowicz}, L.~M., \& {Hawley}, S.~L. 2009, \aj, 137, 3297

\bibitem[{{West} {et~al.}(2008){West}, {Hawley}, {Bochanski}, {Covey}, {Reid},
  {Dhital}, {Hilton}, \& {Masuda}}]{2008AJ....135..785W}
{West}, A.~A., {Hawley}, S.~L., {Bochanski}, J.~J., {Covey}, K.~R., {Reid},
  I.~N., {Dhital}, S., {Hilton}, E.~J., \& {Masuda}, M. 2008, \aj, 135, 785

\end{thebibliography}

\clearpage

\clearpage


\clearpage

\end{document}